\documentclass[aps,floatfix,superscriptaddress,preprintnumbers,showpacs,showkeys]{revtex4}
\usepackage{amssymb}
\usepackage{amsmath}
\usepackage{amsfonts}
\usepackage{epsfig}
\usepackage{graphicx}
\usepackage{tabularx}
\newcommand{\seq}{\begin{subequations}}
\newcommand{\sen}{\end{subequations}}
\newcommand{\eq}{\begin{eqnarray}}
\newcommand{\en}{\end{eqnarray}}

\def\shiftdown#1{#1\llap{\lower.04ex\hbox{#1}}}

\begin{document}

\title{Search for the glueball content of hadrons in $\gamma p$ 
interactions at GlueX} 

\author{Thomas Gutsche} 
\affiliation{Institut f\"ur Theoretische Physik,
Universit\"at T\"ubingen,
Kepler Center for Astro and Particle Physics,
Auf der Morgenstelle 14, D-72076 T\"ubingen, Germany}
\author{Serguei Kuleshov}
\affiliation{Departamento de F\'\i sica y Centro Cient\'\i fico 
Tecnol\'ogico de Valpara\'\i so (CCTVal), Universidad T\'ecnica
Federico Santa Mar\'\i a, Casilla 110-V, Valpara\'\i so, Chile}
\author{Valery E. Lyubovitskij}
\affiliation{Institut f\"ur Theoretische Physik,
Universit\"at T\"ubingen,
Kepler Center for Astro and Particle Physics,
Auf der Morgenstelle 14, D-72076 T\"ubingen, Germany}
\affiliation{Department of Physics, Tomsk State University, 634050 Tomsk, 
Russia}
\affiliation{Laboratory of Particle Physics, Mathematical Physics Department,
Tomsk Polytechnic University, Lenin Avenue 30, 634050 Tomsk, Russia}
\author{Igor T. Obukhovsky}
\affiliation{Institute of Nuclear Physics, Moscow
State University, 119991 Moscow, Russia} 

\today

\begin{abstract}

We suggest a theoretical framework for the description of 
double photon and proton-antiproton photoproduction off the proton 
$\gamma + p \to 2\gamma + p$ and $\gamma + p \to \bar p p + p$  
which takes into account the contribution of the scalar mesons 
$f_0(1370)$, $f_0(1500)$ and $f_0(1710)$. These scalars are considered 
as mixed states of a glueball and nonstrange and strange quarkonia.
Our framework is based on the use of effective hadronic Lagrangians
that phenomenologically take into account two-gluon exchange effects 
governing the $\gamma + p \to 2\gamma + p$ and $\gamma + p \to \bar p p + p$ 
processes. Present results can be used to guide the possible search for
these reactions by the GlueX Collaboration at JLab.

\end{abstract}

\pacs{12.39.Mk,13.60.Fz,14.20.Dh,14.40.Be}

\keywords{hadron structure, scalar mesons, 
glueball and strange content of hadrons, 
phenomenological Lagrangians} 

\maketitle 

\section{Introduction}

The lowest-lying glueballs with quantum numbers $J^{PC} = 0^{++}, 2^{++}$ might 
give significant contributions to hadronic production 
cross sections. In this respect useful reactions could be 
a proton-antiproton pair and photon photoproduction off a proton 
target in exclusive experiments (e.g., the GlueX experiment 
at JLab~\cite{GlueX}) set up for the production of states with positive 
$P$ and $C$ parity, e.g. in two-particle channels with 
$J^{PC} = 0^{++}, 2^{++}$: 
\eq 
\gamma + p \to p+G\to p+(2\gamma,\,2\eta,\,
2\eta^\prime,\,\pi\pi,\,\bar KK,\,\bar pp, \ldots ) \,. 
\en 
When focusing on events without mesons in the final state, 
the detection of the glueball or a glueball component in a hadron 
is significantly simplified.
In this case only the two-body glueball decay channels 
$G \to 2\gamma$ and $G \to \bar p p$ will be present. 
One should stress that the mode $G \to 2 \gamma$ is forbidden when 
the initial state is pure glueball 
and this process can only proceed via mixing of the glueball with nonstrange 
and strange quarkonia components~\cite{Giacosa:2005qr,Chatzis:2011qz}. 
Here the $\bar p p$ pair is in a definite partial wave: 
$^3P_0$ or $^3P_2$ in correspondence with the quantum numbers of the glueball 
$J^{PC} = 0^{++}$ or $ 2^{++}$. 

Due to the constraints mentioned above, the mechanism of the reaction 
\eq 
\gamma + p \to p + G \to p + 2\gamma
\en 
or 
\eq 
\gamma + p  \to p + G \to p + \bar p p
\en 
proceeding through the intermediate glueball $G$ could correspond 
to the search for the glueball content of the scalar $f_0$ and 
tensor $f_2$ mesons.  

\begin{figure}
\begin{center}
\vspace*{1cm}
\epsfig{figure=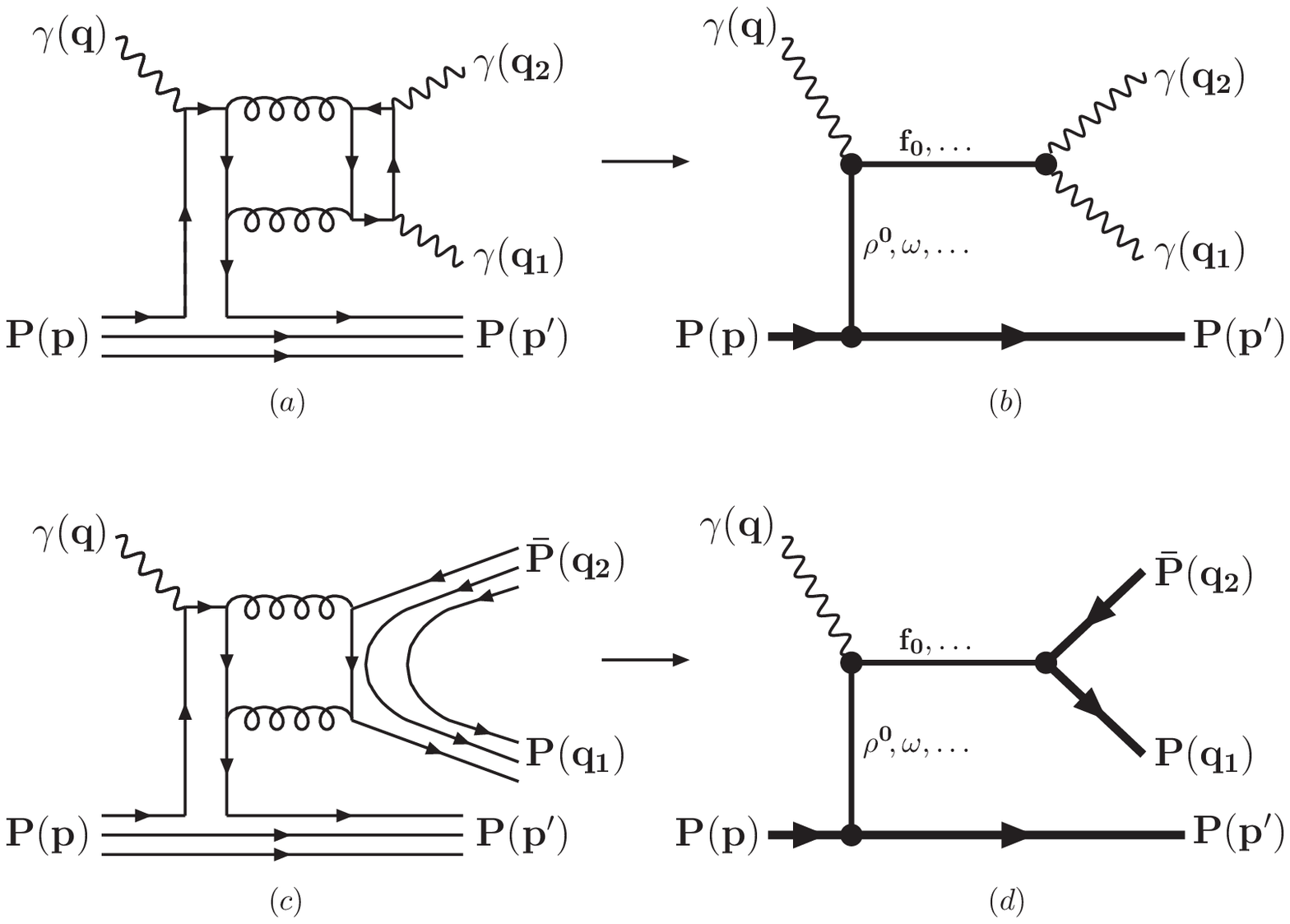,scale=.7}
\caption{QCD $\to$ Effective hadronic Lagrangian correspondence for 
$\gamma p$ reactions. 
\label{fig:fig1}}
\end{center}
\end{figure}

Note that the initial photon has the set of quantum numbers $J^{PC} = 1^{--}$;  
therefore, it could excite resonances with $J^{PC} = 0^{++}$ 
and $2^{++}$ only if it is absorbed by a hadron with 
the same quantum numbers, i.e., vector mesons, $\rho^0$ and 
$\omega$ from the meson cloud of the proton target. 
The Feynman diagrams corresponding to these processes are shown 
in Figs.~\ref{fig:fig1}(a) and~\ref{fig:fig1}(c). 
The QCD diagrams in the left panel of Fig.~\ref{fig:fig1} can effectively be 
understood in terms of diagrams on the hadronic level as displayed
in the right panel [Figs.~\ref{fig:fig1}(b) and~\ref{fig:fig1}(d)].   
In the present manuscript we will concentrate on the analysis of the role 
of scalar mesons. The contribution of tensor mesons will be studied in a 
forthcoming paper. 

The diagrams contributing to the photoproduction of $2\gamma$ and $\bar p p$ 
in the right panel can be calculated on the hadronic level using effective 
Lagrangians including photons~\cite{Giacosa:2005qr}-\cite{Obukhovsky:2009th}. 
These methods were proposed and successfully applied 
in Refs.~\cite{Giacosa:2005qr}-\cite{Gutsche:2008qq} 
for the description of the mass spectrum and decay rates 
of scalar, pseudoscalar, and tensor mesons (including radially 
excited states) and three-body decays of $J/\psi$. 
In particular, our approach is based on the hypothesis of a mixed structure of 
the scalar $f_0(1370)$, $f_0(1500)$, and $f_0(1710)$ mesons, which 
involve the nonstrange and strange quarkonia and glueball 
components\cite{Giacosa:2005qr}-\cite{Chatzis:2011qz}.  
The use of such  effective Lagrangians for light hadrons
composed of light $(u,d,s)$ quarks 
in the energy region up to a few GeV is well justified --- it is 
generalized chiral perturbation theory involving chiral fields, 
low-lying baryons and their resonances. 

Note that taking into account vector meson ($\rho$ and $\omega$) 
exchange does not give a precise description of the amplitudes 
under study. Indeed one should include the Regge poles that will dominate 
in the physical amplitudes. As is well-known, Regge phenomenology is
successfully used in the description of new highly excited meson 
and baryon -precision data on photo- and electroproduction of mesons of 
nucleons and nuclei~\cite{Donnachie:2015jaa}-\cite{daSilva:2013yka} at energies 
$\sqrt{s} \ge 2 - 3$ GeV, allowing the emission of new highly excited 
meson and baryon resonances. For the present kinematical regime,  
Regge phenomenology, which is justified for asymptotically high energies 
$s \gg m_N^2$, allows for a description of photo- and electroproduction 
on a qualitative level. By fitting free parameters, such as little-known 
coupling constants and form factors, one can also obtain a quantitative 
description of data. 

In the photoproduction of mesons with positive 
$P$- and $C$-parity, which could mix with the possibly existing low-lying 
glueball with $J^{PC} = 0^{++}$, the differential cross section is defined by 
exchange diagrams with mesons having negative charge parity and $P = \pm$. 
Note, that because low-lying scalar and pseudoscalar mesons with 
$J^{PC} = 0^{\pm -}$ do not exist, the leading contribution 
to the matrix elements gives exchange by vector and axial-vector mesons with 
$J^{PC} = 1^{\pm -}$. It is sufficient that a Reggeization of these meson 
exchanges leads to an amplitude qualitatively different from the corresponding 
Born amplitude in Born approximation (Feynman diagram). 
In our approach we consider specific processes for which we make an assumption 
that the $\rho$, $\omega$ 
Regge amplitude has a pole at $t \simeq - 0.6$ 
GeV$^2$; i.e., in this range of values of the $t$ variable it is strongly 
suppressed in comparison with the Born amplitude. Note that such an assumption 
is not unique. For instance, in Ref.~\cite{Mathieu:2015gxa} in the consideration of 
pion-nucleon elastic scattering, there are two amplitudes
involving the $\rho$ Regge poles. One has a zero for $t \sim - 0.1$GeV$^2$. 

Reggeization of the diagrams corresponding to the $\rho$ and $\omega$ meson 
exchange effectively leads to a replacement of vector mesons poles in the 
propagators by poles lying on the linear angular-momentum $J = \alpha(t)$ 
Regge trajectories $\alpha_V(t)=\alpha_{0V}+\alpha^\prime_Vt$ (see details 
below). Axial-vector meson exchange is Reggeized in analogy, but it does not 
change the final result very much because in the reaction under consideration 
axial-vector exchange does not interfere with the vector one. A sizable 
influence on the final result comes from the uncertainty related to poor 
knowledge of meson-nucleon coupling constants. 
In the present manuscript we will consider two scenarios for the choice of the set 
of vector couplings: 
1) Variant I, the standard ``soft set'' of vector couplings and 
2) Variant II, the``hard set'' of vector coupling constants 
usually used in the Regge approach to photo- and 
electroproduction of mesons~\cite{Donnachie:2015jaa}-\cite{daSilva:2013yka}. 
In both cases the parameters of the Regge trajectories will also be different.  
In both scenarios we will use the same specific mixing scheme of scalar mesons 
obtained in Ref.~\cite{Chatzis:2011qz} from the analysis of $J/\psi$ decays. 

The paper is organized as follows. In Sec.~II, we derive the 
formalism for the description of photoproduction off the proton target. 
We present the phenomenological hadronic Lagrangian, including photons. 
Then we calculate the matrix elements for the process under study. 
In Sec.~III we discuss our numerical results for the differential and
double-differential 
cross sections for two-photon and nucleon-antinucleon photoproduction. 
Finally, we give a brief conclusion. 
 
\section{Formalism} 

We start with the definition of the phenomenological Lagrangians. We indicate 
the terms that are relevant for the description of photon-proton collisions,  
including the resonance contributions of the scalar mesons $f_0(1370)$, 
$f_0(1500)$, and $f_0(1710)$. 
The full interaction Lagrangian~\cite{Giacosa:2005qr,Chatzis:2011qz}, 
relevant for the  description of the photoproduction processes  
$\gamma + p \to 2\gamma + p$ and $\gamma + p \to \bar p p + p$ 
is given by a sum of interaction Lagrangians
\eq 
{\cal L}_{\rm full}(x) = 
{\cal L}_{VNN}(x) + 
{\cal L}_{f_0NN}(x) + 
{\cal L}_{f_0\gamma\gamma}(x) + 
{\cal L}_{f_0V\gamma}(x)
\en 
where 
\eq 
{\cal L}_{VNN}(x) &=& 
g_{\rho NN} \, \bar N(x) \gamma^\mu 
\vec{\rho}_\mu(x) \vec{\tau} N(x) 
+ \frac{f_{\rho NN}}{4 M_N} \, \bar N(x) \sigma^{\mu\nu}  
\vec{R}_{\mu\nu}(x) \vec{\tau} N(x) \nonumber\\ 
&+& 
g_{\omega NN} \, \bar N(x) \gamma^\mu 
\omega_\mu(x) N(x) 
+ \frac{f_{\omega NN}}{4 M_N} \, \bar N(x) \sigma^{\mu\nu}  
W_{\mu\nu}(x) N(x) \,, \\
{\cal L}_{f_0NN}(x) &=& F_{\mu\nu}(x) \, F^{\mu\nu}(x) \, 
\sum\limits_i \, g_{f_iNN} \, f_i(x) \,,  \\
{\cal L}_{f_0\gamma\gamma}(x) &=& \frac{e^2}{4} \, 
F_{\mu\nu}(x) \, F^{\mu\nu}(x) \, 
\sum\limits_i \, g_{f_i\gamma\gamma} \, f_i(x) \, 
 \,, \\
{\cal L}_{f_0V\gamma}(x) &=& \frac{e}{2} \, F_{\mu\nu}(x) \,  
\sum\limits_i \, \biggl[g_{f_i\rho\gamma} \, f_i(x) \, R^{0, \mu\nu}(x) + 
g_{f_i\omega\gamma} \, f_i(x) \, W^{\mu\nu}(x)\biggr] 
\,.
\en 
Here we introduce the following notation: 
$f_i$ is the set of scalar mesons --- 
$f_1 = f_0(1370)$, $f_2 = f_0(1500)$, and $f_3 = f_0(1710)$; 
$F_{\mu\nu} = \partial_\mu A_\nu - \partial_\nu A_\mu$, 
$R_{\mu\nu}^i = \partial_\mu \rho_\nu^i - \partial_\nu \rho_\mu^i$ 
and 
$W_{\mu\nu} = \partial_\mu \omega_\nu - \partial_\nu \omega_\mu$ 
are the stress tensors of the electromagnetic field and  
$\rho$ and $\omega$ mesons, respectively.  

The scalar fields are considered as mixed states of the glueball $G$ 
and nonstrange $N$ and strange $S$ 
quarkonia~\cite{Giacosa:2005qr,Chatzis:2011qz} (see 
more details in the Appendix and the review~\cite{Klempt:2007cp} 
for a discussion on different interpretations of these states):
\eq 
f_i = B_{i1} {\cal N} + B_{i2} G + B_{i3} S \, .  
\en 
The $B_{ij}$ are the elements of the mixing matrix rotating 
bare states $({\cal N}, G, S)$ into the physical scalar mesons 
$(f_0(1370), f_0(1500), f_0(1710))$. 
In Refs.~\cite{Giacosa:2005qr,Chatzis:2011qz} 
we studied in detail different scenarios for the mixing of 
${\cal N}$, $G$, and $S$ states. Here we proceed with the scenario 
fixed in Ref.~\cite{Chatzis:2011qz} from a full analysis 
of strong $f_0$ decays and radiative decays of the $J/\psi$ with
the scalars in the final state: 
\eq\label{Bmatrix} 
B=\left(
\begin{array}{lll}
   0.75 & 0.60    & 0.26    \\
$-0.59$ & 0.80    & $-0.14$ \\
$-0.29$ & $-0.05$ & 0.95    \\ 
\end{array}
\right) \,.
\en
The coupling constants involving scalar mesons are given 
in terms of the matrix elements $B_{ij}$ and the effective couplings 
$c_e^s$, $c_e^g$, $c_f^s$, and $c_f^g$ 
of~\cite{Chatzis:2011qz}:  
\eq 
g_{f_i\gamma\gamma} &=& 
  \frac{80}{9 \sqrt{2}} \, c_e^s \, B_{i1} 
+ \frac{32}{\sqrt{3}}   \, c_e^g \, B_{i2} 
+ \frac{16}{9} \, c_e^s \, B_{i3} \,, \nonumber\\
g_{f_i\rho\gamma} &=& 3 g_{f_i\omega\gamma} 
\ = \ B_{i1} \, c_f^s + B_{i2} \, \sqrt{\frac{2}{3}} \, c_f^g \,.
\en   
The effective couplings 
$c_e^s = 0.056$ GeV$^{-1}$, $c_e^g = 0.003$ GeV$^{-1}$,  
$c_f^s = 1.592$ GeV$^{-1}$, and $c_f^g = 0.078$ GeV$^{-1}$ are 
fixed from data involving the scalar mesons $f_i$. 
In case of the $f_iNN$ couplings we suppose that they are dominated by the 
coupling of the nonstrange component to the nucleon,  
\eq
g_{f_iNN} \simeq B_{i1} \, g_{{\cal N}NN} \,.
\en
The coupling $g_{{\cal N}NN}$ can be identified with the 
coupling of the nonstrange scalar $\sigma$ meson to nucleons,  
\eq\label{gsigmaNNcoupling}  
g_{{\cal N}NN} =  g_{\sigma NN} \simeq 5 \,,  
\en  
which plays an important role in phenomenological approaches 
to the nucleon-nucleon potential generated by
meson exchange~\cite{Machleidt:2000ge}. 
 
Next we write down the standard expressions for the invariant amplitudes 
${\cal M}_{inv}^{\gamma\gamma}$ and ${\cal M}_{inv}^{p\bar p}$ 
sketched in Figs.~\ref{fig:fig1}(b) and (d): 
\eq
{\cal M}_{\rm inv}^{\gamma\gamma}&=&
\sum_{V}\sum_{i}
\, G_{eff}^{\gamma\gamma}(i,V)
\,\bar u(p') 
\biggl[ \gamma^\mu G_V + \frac{i\sigma^{\mu\nu} k_\nu}{2M_N} F_V \biggr] u(p) 
\, \frac{- g_{\mu\alpha} + k_\mu k_\alpha/M_V^2}{M_V^2 - k^2 + i \Gamma_V M_V} 
\nonumber\\
&\times& (kq \, g^{\alpha\beta} - k^\beta q^\alpha) 
\, \frac{1}{M_{f_i}^2 - l^2 + i \Gamma_{f_i} M_{f_i}}\,
[q_1^\sigma q_2^\rho-g^{\rho\sigma}q_1 q_2]
\, \epsilon_\beta(q) \, \epsilon_\rho^\ast (q_1) \, \epsilon_\sigma^\ast(q_2)
\en
and
\eq
{\cal M}_{\rm inv}^{p\bar p}&=&
\sum_{V}\sum_{i} \, 
G_{eff}^{p\bar p}(i,V)\,\bar u(p') 
\biggl[ \gamma^\mu G_V + \frac{i\sigma^{\mu\nu} k_\nu}{2M_N} F_V \biggr] u(p) 
\, \frac{- g_{\mu\alpha} + k_\mu k_\alpha/M_V^2}{M_V^2 - k^2 + i \Gamma_V M_V} 
\nonumber\\
&\times& (kq \, g^{\alpha\beta} - k^\beta q^\alpha) 
\, \frac{1}{M_{f_i}^2 - l^2 + i \Gamma_{f_i} M_{f_i}} 
\, \epsilon_\beta(q) \, \bar u(q_1) \, v(q_2)\,.
\en 
The momenta $p$ and $p'$ are of the initial and final proton;  
$q$ is the photon momentum in the initial state;  
$q_1$ and $q_2$ are the momenta of two photons or 
the nucleon and antinucleon in the final state; and  
$k$ and $l$ ($l=k+q$) are the intermediate vector and scalar meson momenta, 
respectively. Here the constants
$G_{eff}^{\gamma\gamma}(i,V)=g_{f_iV\gamma}g_{f_i\gamma\gamma}$ and
$G_{eff}^{p\bar p}(i,V)=g_{f_iV\gamma}g_{f_i NN}$ are 
the effective couplings for $\gamma\gamma$ and $p\bar p$ photoproduction. 

The expression for the twofold differential cross section 
of the double photon production,  
\eq 
\frac{d\sigma_{\gamma\gamma}}{ds_2 dt_1} = \frac{\alpha^3}{4} \, 
\frac{1}{(s-m_N^2)^2} \, \overline{\vert{\cal M}_{\gamma\gamma}\vert^2} 
\en
implies the summation over contributions from intermediate mesons, 
$V=\rho,\,\omega$ and $f_i,\,i=1,2,3$, to the amplitude. The
spin-averaged square of the amplitude is given by
\eq
\overline{\vert{\cal M}_{\gamma\gamma}\vert^2}\,&=& \,
\frac{1}{4} \, \sum_{\lambda_\gamma=\pm 1} \,\frac{1}{2} \, \sum_{s_z=\pm 1/2}
\biggl\vert \sum_{V}\sum_{i}
{\cal M}^{\gamma\gamma}_{\rm inv}\biggr\vert^2 \nonumber\\
&=&- \frac{s_2^2}{2} \, 
\sum_{V,V^\prime}\, \sum_{i,j=1}^3 \, G_{fV\gamma}(i,j;V,V^\prime)
T_{VNN}(V,V^\prime;s,s_2,t_1) \, D_V(V,V^\prime ;t_1) \,  D_f(i,j,s_2) \,,
\en 
where $s = (p + q)^2$, $s_2 = (q_1+q_2)^2$, and $t_1 = (p'-p)^2$ are
Mandelstam variables, corresponding to the total energy and transverse momentum 
squared to the two-photon pair and to the vector meson. The values of $s_2$
and $t_1$ are kinematically constrained to the intervals
\eq
0\le\sqrt{s_2}\le\sqrt{s}-m_N,\quad
t_{min}\le t_1 \le t_{max},
\en
where 
\eq
t_{max/min}=2m_N^2-\frac{1}{2s}[(s+m_N^2)(s-s_2+m_N^2)\mp (s-m_N^2)
\lambda^{1/2}(s,s_2,m_N^2)],
\en
and $\alpha = 1/137.036$ is the fine-structure constant.  

The vertex functions 
$G_{fV\gamma}(i,j;V,V^\prime)$, $T_{VNN}(V,V^\prime;s,s_2,t_1)$, 
$D_V(V,V^\prime;t_1)$, and $D_f(i,j,s_2)$ are defined 
as 
\eq 
G_{fV\gamma}(i,j;V,V^\prime)&=&g_{f_i\gamma\gamma}g_{f_j\gamma\gamma}
g_{f_i V\gamma} g_{f_j V^\prime\gamma}\,,\nonumber\\
T_{VNN}(V,V^\prime;s,s_2,t_1) &=& 
(G_V + F_V) \, (G_{V^\prime} + F_{V^\prime}) \, 
\frac{t_1 (t_1 - s_2)^2}{4}+\biggl(G_V G_{V^\prime} \,-\, 
\frac{t_1}{4 m_N^2} F_V F_{V^\prime}\biggr) \nonumber\\ 
&\times&
\biggl( \frac{m_N^2}{4} \, (t_1-s_2)^2 \,+\, 
\frac{t_1}{4} \, (s-m_N^2) \, (s-m_N^2+t_1-s_2)\biggr)\,,\nonumber\\ 
D_V(V,V^\prime;t_1)&=&\frac{(t_1-M_V^2)(t_1-M_{V^\prime}^2)+
M_V\Gamma_V M_{V^\prime}\Gamma_{V^\prime}}
{[{(t_1-M_V^2)}^2+M_V^2\Gamma_V^2][{(t_1-M_{V^\prime}^2)}^2+
M_{V^\prime}^2\Gamma_{V^\prime}^2]}\,,\nonumber\\
D_f(i,j;s_2)&=&\frac{(s_2-M_{f_i}^2)(s_2-M_{f_j}^2)
+M_{f_i}\Gamma_{f_i}M_{f_j}\Gamma_{f_j}}
{[{(s_2-M_{f_i}^2)}^2 +M_{f_i}^2\Gamma_{f_i}^2]
 [{(s_2-M_{f_j}^2)}^2 +M_{f_j}^2\Gamma_{f_j}^2]} \,,
\en 
where the indices $i,j = 1,2,3$ number the scalar $f_0(1370)$, 
$f_0(1500)$, and $f_0(1710)$ mesons and indices $V,V^\prime$ correspond the 
$\rho$ and $\omega$ vector mesons;  
$g_{VNN}$ and $f_{VNN}$ are the Dirac and Pauli coupling constants 
of vector mesons to nucleons, respectively; and $g_{fV\gamma}$ are the couplings 
of vector and scalar mesons with photons. 

Note that for GlueX energies the leading contribution to the matrix element 
of $2\gamma$ production scales as $G_V G_{V'} t_1 s^2$, 
where the factor $t_1 s^2$ comes from the helicity flip 
in the $\gamma-f_0$ vertex squared.

The meson parameters can be grouped as follows::
 
(1) Parameters involving masses, widths and couplings of scalar mesons 
have been fixed from data with~\cite{Chatzis:2011qz}  
\eq\label{parameter_set1} 
& &M_{f_1} = 1.432\, \mathrm{GeV}\,,\quad 
   M_{f_2} = 1.510\, \mathrm{GeV}\,,\quad 
   M_{f_3} = 1.720\, \mathrm{GeV}\,, \nonumber\\
& &\Gamma_{f_1} = 0.350\, \mathrm{GeV}\,,\quad 
   \Gamma_{f_2} = 0.109\, \mathrm{GeV}\,,\quad 
   \Gamma_{f_3} = 0.135\, \mathrm{GeV}\,, \nonumber\\
& &g_{f_1\gamma\gamma} =  0.30\, \mathrm{GeV}^{-1} \,,\quad 
   g_{f_2\gamma\gamma} = -0.21\, \mathrm{GeV}^{-1} \,,\quad 
   g_{f_3\gamma\gamma} = -0.01\, \mathrm{GeV}^{-1} \,,\nonumber\\
& &g_{f_1 V\gamma} =  1.24\, \mathrm{GeV}^{-1} \,,\quad 
   g_{f_2 V\gamma} = -0.90\, \mathrm{GeV}^{-1} \,,\quad 
   g_{f_3 V\gamma} = -0.47\, \mathrm{GeV}^{-1} \,,\nonumber\\
& &g_{f_1 NN} =    3.75\,,\quad  
   g_{f_2 NN} =  - 2.95\,,\quad  
   g_{f_3 NN} =  - 1.45\,. 
\en
(2) Parameters involving masses and widths of vector mesons taken are
from the PDG compilation~\cite{Agashe:2014kda}  
\eq 
& &M_\rho = 0.7755\, \mathrm{GeV}\,,\quad 
   M_\omega = 0.7866\, \mathrm{GeV}\,, \nonumber\\
& &\Gamma_\rho = 0.149\, \mathrm{GeV}\,,\quad 
   \Gamma_\omega = 0.0085\, \mathrm{GeV}\,, \nonumber\\
& &G_\rho = 2.3 \,, \quad 
   G_\omega = 3 G_\rho \,,\nonumber\\ 
& &F_\rho = 3.66 G_\rho \,, \quad 
   F_\omega = -0.07 G_\omega \,. 
\en  

For the process $\gamma p\to p+p\bar p$ one has correspondingly
\eq
\frac{d\sigma_{p\bar p}}{ds_2\,dt}=\frac{\alpha}{32\pi^2(s-m_N^2)}
\overline{\vert{\cal M}_{p\bar p}\vert^2}\,,\qquad 
2m_N\le\sqrt{s_2}\le\sqrt{s}-m_N,\quad t_{min}\le t_1 \le t_{max}\,,
\en
where 
\eq
\overline{\vert{\cal M}_{p\bar p}\vert^2}\,&=& \,
\frac{1}{2} \, \sum_{\lambda_\gamma=\pm 1} \,\frac{1}{2} \, \sum_{s_z=\pm 1/2}
\biggl\vert \sum_{V}\sum_{i}{\cal M}^{p\bar p}_{inv}\biggr\vert^2 
\nonumber\\
&=&- (s_2-4m_N^2) \, 
\sum_{V,V^\prime}\, \sum_{i,j=1}^3 \, G_{fVN}(i,j;V,V^\prime)
T_{VNN}(V,V^\prime;s,s_2,t_1) \, D_V(V,V^\prime ;t_1) \,  D_f(i,j,s_2) \,
\en
and 
\eq 
G_{fVN}(i,j;V,V^\prime)= g_{f_i N\bar N} \, g_{f_j N\bar N} \, 
g_{f_i V\gamma} \, g_{f_j V^\prime\gamma}\,.
\en 

As we stressed already in the Introduction, the vector meson exchanges in the 
$t$ channel are Reggeizied. In particular, Reggeization of the diagrams 
corresponding to the $\rho$ and $\omega$ meson exchange effectively 
leads to the replacement of Feynman propagators by contributions 
of poles lying on the angular-momentum $J = \alpha(t)$ 
linear Regge trajectories $\alpha_V(t)=\alpha_{0V}+\alpha^\prime_Vt$ as   
\eq 
\frac{1}{t-m_V^2}\,\to\,\left(\frac{s}{s_0}\right)^{\alpha_V(t)-1}
(-\alpha_V^\prime)\,\Gamma(1\!-\!\alpha_V(t))\frac{-1+e^{i\pi\alpha_V(t)}}{2}.
\label{regge}
\en 
As before, in the present manuscript we consider two scenarios for the choice 
of the set of vector couplings 

\noindent
(1) Variant I, the standard ``soft set'' of 
$G_\rho,\,G_\omega,\,F_\rho,\,F_\omega$ couplings and

\noindent 
(2) Variant II, the ``hard set'' of vector coupling constants 
$G_\rho=3.4$  , $G_\omega=15$, $F_\rho=20.7$, and $F_\omega=0$ usually used in 
the Regge approach to photo- and electroproduction of 
mesons~\cite{Donnachie:2015jaa,daSilva:2013yka}. 

\noindent
The parameters of the Regge trajectories will also be different in both cases: 
$\alpha_{0\rho}=0.53$,
$\alpha_\rho^\prime =0.85$\,GeV$^{-2}$,  $\alpha_{0\omega}= 0.4$,
$\alpha_\omega^\prime =0.85$\,GeV$^{-2}$ in the case of Variant I and 
$\alpha_{0\rho}=0.55$,
$\alpha_\rho^\prime =0.8$\,GeV$^{-2}$, and $\alpha_{0\omega}= 0.44$,
$\alpha_\omega^\prime =0.9$\,GeV$^{-2}$ in the case of Variant~II. 
In both scenarios we will use the same specific mixing scheme of scalar mesons 
obtained in Ref.~\cite{Chatzis:2011qz}. The couplings of scalar mesons
to vector mesons and photons [see Eq.~(\ref{parameter_set1})] have been
fixed in Ref.~\cite{Chatzis:2011qz} from the analysis of $J/\psi$ decays.

\section{Results}

In Fig.~2 we present the results of our calculation for the differential cross 
section $d\sigma/dt$ (as an integral of the double differential cross section 
$d\sigma_{\gamma\gamma}/dtds_2$ of $2\gamma$ photoproduction  over the Mandelstam
variable $ds_2$) for two values for the energy of the initial photon: 
$E_\gamma=\,$5 and 9 GeV. We compare results for these two variants 
of the coupling constants (``soft'' and ``hard''). Also for a 
comparison we present results of calculations using Feynman diagrams with 
vector meson exchange and taking into account its finite width. 

The obtained results can be useful for an estimate of a signal from a mixed 
($f_0+G$) scalar meson in the $2\gamma$ channel. In this case the 
contributions of other subprocesses into $2\gamma$ photoproduction [for 
example via photoproduction of $\pi^0$, $f_0/a_0(980)$ or other mesons of 
positive charge parity] we consider as background, which can be estimated using 
known calculations of meson 
photoproduction~\cite{Donnachie:2015jaa}-\cite{daSilva:2013yka}.  
In Fig.3 we display the double-differential cross section 
$d\sigma_{\gamma\gamma}/dtds_2$ at energies 5 and 9 GeV for our basic scenario 
(Variant I). 

In Fig.~4 the results on $d\sigma/dt$ and 
$d\sigma/dtds_2$ for $p\bar p$ photoproduction at energy $E_\gamma=\,$9 
GeV are also displayed  for Variant I.
Due to the threshold of $p\bar p$ photoproduction, which lies above 
the resonance peaks related 
to the production of the scalar mesons $f_0(1370)$, 
$f_0(1500)$, and $f_0(1710)$, the differential cross section of 
the $\gamma + p \to p + G \to p + p\bar p$ reaction has no resonance 
structure over the physical range of the $s_2$ variable. 
In the two-photon photoproduction corresponding resonance peaks 
play a very important role as seen from the direct 
comparison of Figs.~2, 3 and 4.

The absolute value of the cross section is defined in our approach by 
the parameters and coupling constants (\ref{Bmatrix})-(\ref{gsigmaNNcoupling}), 
which have been fixed before in the analysis of data on scalar 
mesons~\cite{Giacosa:2005qr,Chatzis:2011qz}. 
The $s_2$ dependence of the 
differential cross section is generated by the contribution of 
the scalar meson propagators and the $Vpp$ vertex. For a model dependence 
of the coupling constants in the $Vpp$
vertex we have considered two variants, 
``soft'' (Variant I) and ``hard'' (Variant II). The hard variant results in
differential cross sections that are several times greater than the soft case. 
Nevertheless both variants predict a magnitude of the differential cross 
section $d\sigma_{\gamma\gamma}/dt$ of about 10 $nb/GeV^2$ or greater in the
region of the peak at -$t\approx$0.1-0.2 GeV$^2$. These values are tentative, 
but they may be used as a guide for further research. 

It should be noted that the Regge pole amplitude is suppressed (because of 
the zero at $t\approx\,$-0.6 GeV$^2$) when compared to the contribution of
the vector-meson exchange diagrams (dotted-dashed curves in Figs. 2 and 4).
Recently~\cite{Donnachie:2015jaa} it was shown that the contribution of the 
Regge cut can considerably compensate this drawback at least in meson
photoproduction for energies of $E_\gamma=\,$2-3 GeV. It seems reasonable  
that the Regge-pole terms should be used only as a rough approximation in the
considered region of small values of $\sqrt{s}\approx\,$2-4 GeV.  
 
It is important to mention scalings of the cross sections
considered in the present manuscript on $s$ and $s_2$ variables. 
The differential cross sections scales at large $s$, $s_2$ as 
${\cal O}(1)$ for two-photon photoproduction and 
${\cal O}(s)$, ${\cal O}(s_2)$ for 
proton-antiproton photoproduction.  

In conclusion, we proposed the approach for the description of
double photon and proton-antiproton photoproduction off the proton, 
which takes into account the contribution of the scalar mesons
$f_0(1370)$, $f_0(1500)$, and $f_0(1710)$. The mesons are considered
as mixed states of a glueball and nonstrange and strange quarkonia.
Our framework is based on the use of effective hadronic Lagrangians 
that take into account two-gluon exchange effects
governing the $\gamma + p \to 2\gamma + p$ and $\gamma + p \to \bar p p + p$
processes. We presented results that can be used for a possible search
for these reactions by the GlueX Collaboration at JLab.

\begin{acknowledgments}

This work was supported
by the German Bundesministerium f\"ur Bildung und Forschung (BMBF)
under Project 05P2015 - ALICE at High Rate (BMBF-FSP 202):
``Jet- and fragmentation processes at ALICE and the parton structure      
of nuclei and structure of heavy hadrons'', 
by the Basal Conicyt No. FB082, 
by Fondecyt (Chile) Grant No. 1140471 and CONICYT (Chile) Grant No. ACT1406,
by Tomsk State University Competitiveness
Improvement Program and the Russian Federation program ``Nauka''
(Contract No. 0.1526.2015, 3854), by the Deutsche Forschungsgemeinschaft
(DFG-Project FA 67/42-1 and GU 267/3-1) and by the Russian Foundation 
for Basic Research (Grant No. RFBR-DFG-a 16-52-12019).

\end{acknowledgments}

\appendix
\section{Mixing of scalar mesons} 

Here we briefly discuss the mixing of the glueball with 
the scalar-isoscalar quarkonia states~\cite{Giacosa:2005qr,Chatzis:2011qz} 
and octet $(S^{8})$ scalar quarkonia states. 

We start with the Lagrangian involving the scalar mesons and glueball, 
\eq\label{L_eff}
{\cal L} = \frac{1}{2}\left\langle
D_{\mu }{\cal S}D^{\mu }{\cal S}-M_{{\cal S}}^{2}{\cal S}^{2}\right\rangle 
+ \frac{1}{2}(\partial _{\mu }G\partial ^{\mu }G-M_{G}^{2}G^{2}) 
\en
where ${\cal S}=\sum_{i=0}^{8}S_{i}\lambda _{i}/\sqrt{2}$. 
Singlet $S^{0}$ and octet $S^{8}$ components are linear combinations 
of nonstrange $N$ and strange $S$ components with the respective flavor content
$[\bar{u}u+\bar{d}d]/\sqrt{2}$ and $\bar{s}s$ 
\eq 
S^{0}
\,=\,\sqrt{2/3}\,N\,+\,\sqrt{1/3}\,S\,,\hspace*{0.25cm}S^{8}\,=\,
\sqrt{1/3}\,N-\,\sqrt{2/3}\,S\,.
\en                                               
In Eq.~(\ref{L_eff}) 
the scalar nonet states have the same mass $M_{{\cal S}},$
corresponding to the flavor and large $N_{c}$ limits. Deviations from this
configuration (i.e., from the large $N_{c}$ limit) are encoded in 
$\,{\cal L}_{{\rm mix}}^{S}$,   
including quarkonia-glueball mixing, leading to different
masses for the scalar mesons (in~\cite{Cirigliano:2003yq} this is explicitly
shown for the scalar nonet, but no glueball is included; the glueball
mixing is also a deviation from large $N_{c},$ since in this limit the
glueball and the quarkonia states decouple).
As a result, the scalar-isoscalar
sector can be described by the most general Klein-Gordon (KG) Lagrangian, 
including mixing among the $N$, $S$, and the bare glueball $G$ 
configurations~\cite{Amsler:1995tu,Lee:1999kv,Close:2001ga,%
Burakovsky:1998zg,Strohmeier-Presicek:1999yv}:
\begin{equation}
{\cal L}_{KG}=-\frac{1}{2}\sum\limits_{\Phi =N,G,S}\,\Phi \lbrack \Box
+M_{\Phi }^{2}]\Phi -fGS-\sqrt{2}frGN-\varepsilon NS\,,  \label{L_mix}
\end{equation}
where $\Box =\partial ^{\mu }\partial _{\mu }$. The parameter $f$ is the
quarkonia-glueball mixing strength, analogous to the parameter $z$ of 
Refs.~\cite{Amsler:1995tu,Lee:1999kv,Close:2001ga,%
Burakovsky:1998zg,Strohmeier-Presicek:1999yv};
$z$ refers to the quantum mechanical case where the mass matrix is linear
in the bare mass terms, $f$ in turn is related to the quadratic
Klein-Gordon case. The connection between $f$ and $z$ discussed in 
Refs.~\cite{Burakovsky:1998zg,Giacosa:2004ug} leads to the approximate 
relation 
$f\simeq 2zM_{G}\,.$ If $r=1$, then the glueball is flavor blind and mixes
only with the quarkonium flavor singlet $S^{0}$ (this is the case in 
Refs.~\cite{Amsler:1995tu,Burakovsky:1998zg,Strohmeier-Presicek:1999yv}); 
a value $r\neq 1$ takes into account a possible deviation from this limit. 
However, as deduced from the analyses of 
Refs.~\cite{Lee:1999kv,Close:2001ga,Giacosa:2004ug} $r$ is believed 
to be close to unity. In the following we will use the limit $r=1$;  
i.e., we restrict to a flavor blind mixing.

The parameter $\varepsilon$ induces a direct mixing between $N$ and $S$
quarkonia states. This effect is neglected 
in~\cite{Amsler:1995tu,Burakovsky:1998zg,%
Close:2001ga,Strohmeier-Presicek:1999yv},
where flavor mixing is considered a higher-order effect. However, a
substantial $N$-$S$ mixing in the scalar sector is the starting point of the
analysis of Refs.~\cite{Minkowski:2002nf,Minkowski:1998mf}. The origin of
quarkonia flavor mixing is, according 
to~\cite{Minkowski:2002nf,Minkowski:1998mf}, 
connected to instantons as in the
pseudoscalar channel, but with opposite sign (see also~\cite{Klempt:1995ku}): 
the mixed physical fields are predicted to be a higher-lying state with
flavor structure $[N\sqrt{2}-S]/\sqrt{3}$ and a lower one with 
$[N+S\sqrt{2}]/\sqrt{3}$. Here we study the case $\varepsilon \neq 0,$ 
more precisely $\varepsilon >0,$ which leads to the same phase structure as in 
Refs.~\cite{Minkowski:2002nf,Minkowski:1998mf}, but the quantitative results 
and interpretation will differ. The physical scalar states $\left\vert
i\right\rangle $ are identified with $i=f_{1}\equiv f_{0}(1370)$, 
$f_{2}\equiv $ $f_{0}(1500)$, and $f_{3}\equiv f_{0}(1710)$, which are set up                                                  
as linear combinations of the bare states: $\left\vert i\right\rangle
=B^{iN}|N\rangle +B^{iG}\left\vert G\right\rangle +B^{iS}\left\vert
S\right\rangle $. The amplitudes $B^{ij}$ are the elements of a matrix $B$   
that diagonalize the mass matrix of bare states including mixing, 
which in turn gives rise to the mass matrix of physical states 
$\Omega ^{\prime }= {\rm diag}\{M_{f_{1}}^{2},M_{f_{2}}^{2},M_{f_{3}}^{2}\}\,.$
In the following we use~\cite{Giacosa:2005qr,Chatzis:2011qz} 
a best fit of the parameters entering 
in Eqs.~(\ref{L_eff}) and (\ref{L_mix}) to the experimental averages 
of masses and decay modes listed by PDG~\cite{Agashe:2014kda}.  
In the present manuscript we use the latest set of the parameters fixed 
in Ref.~\cite{Chatzis:2011qz} with accuracy $\chi_{\rm tot}^2 = 19.82$ 
[see Eq.~(\ref{parameter_set1})]. 

\label{AppendixA}

\newpage

\begin{figure}
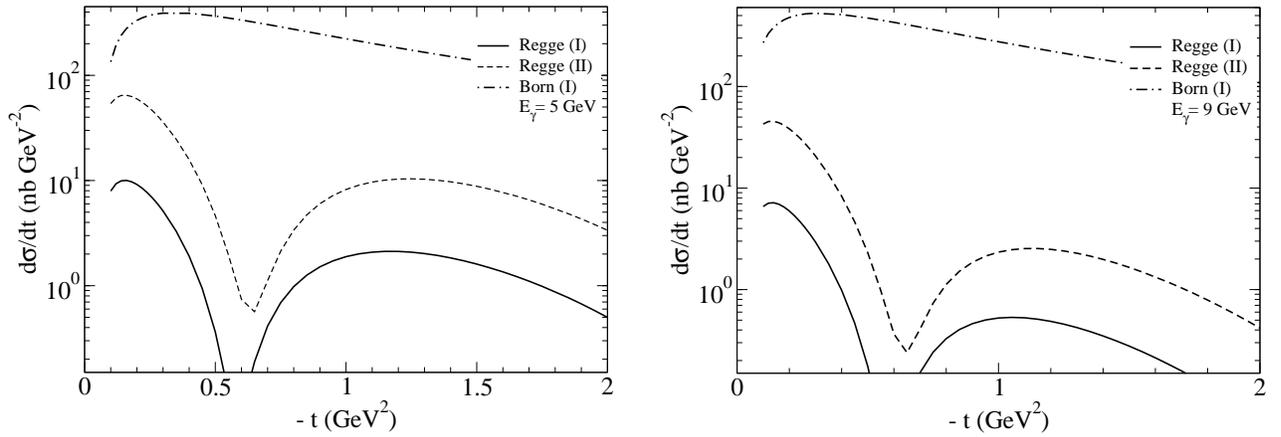

\vspace{5mm}
\begin{center}
\epsfig{figure=fig2a.eps,width=0.45\textwidth}\hspace{5mm}
\epsfig{figure=fig2b.eps,width=0.45\textwidth}
\caption{Differential cross section $d\sigma/dt_1$ (in units of nb/GeV$^2$)
for $\gamma\gamma$ photoproduction off the proton in the model based on the
mixed structure of the scalar $f_0(1370)$, $f_0(1500)$, and $f_0(1710)$
mesons. The initial photon energies are $E_\gamma=\,$5 GeV (left panel) 
and 9 GeV (right panel). Solid and short-dashed curves: Theoretical 
predictions in the glueball model with the Reggeizided vector meson exchange for
soft (Variant I) and hard (Variant II) $VNN$ coupling constants, respectively. 
Dotted-dashed curves: Results in the same
model, but without Reggeization of the vector meson exchange.
\label{fig:fig2}}
\end{center}
\end{figure}

\begin{figure}
\vspace{5mm}
\begin{center}
\epsfig{figure=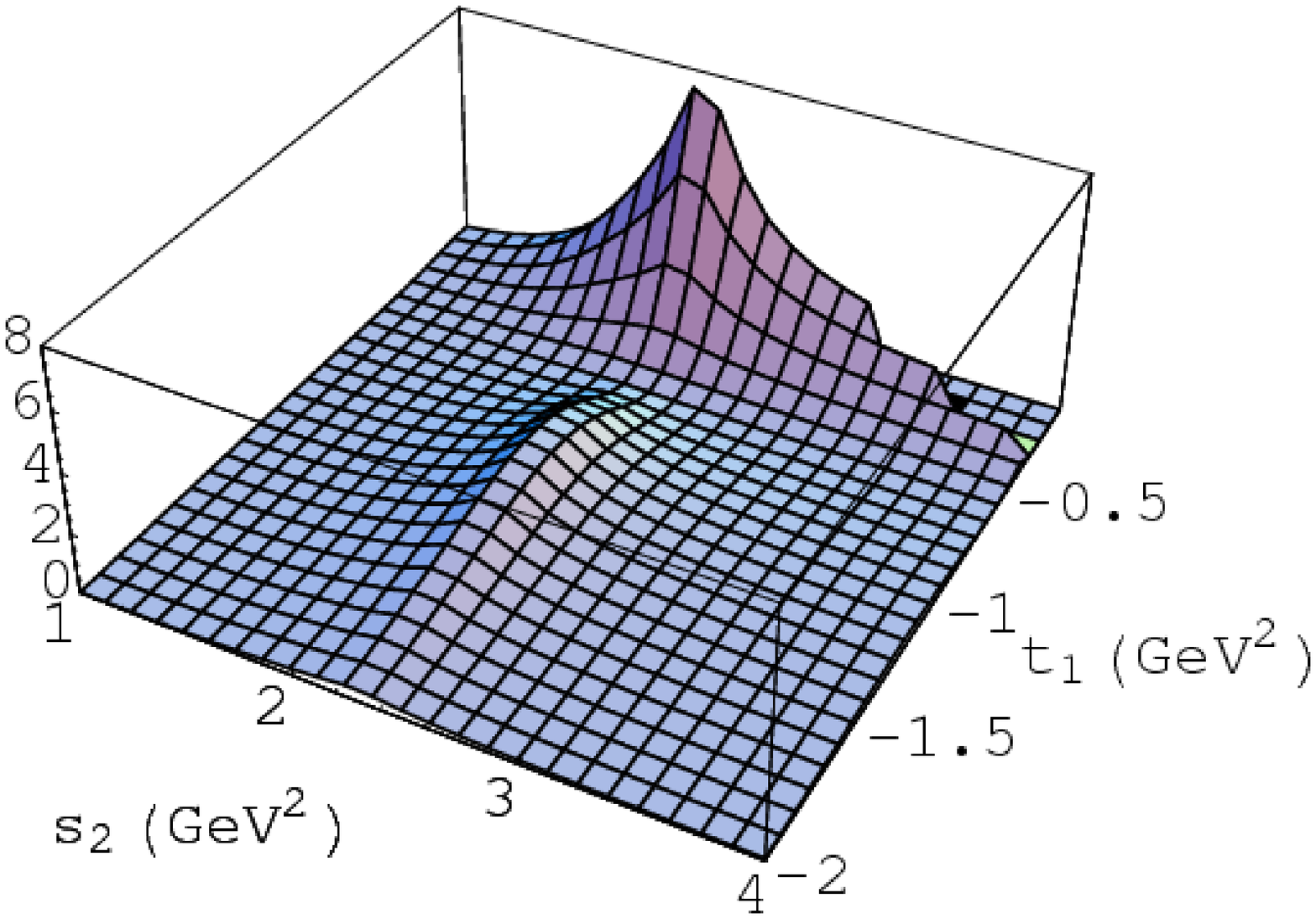,width=0.45\textwidth}\hspace{5mm}
\epsfig{figure=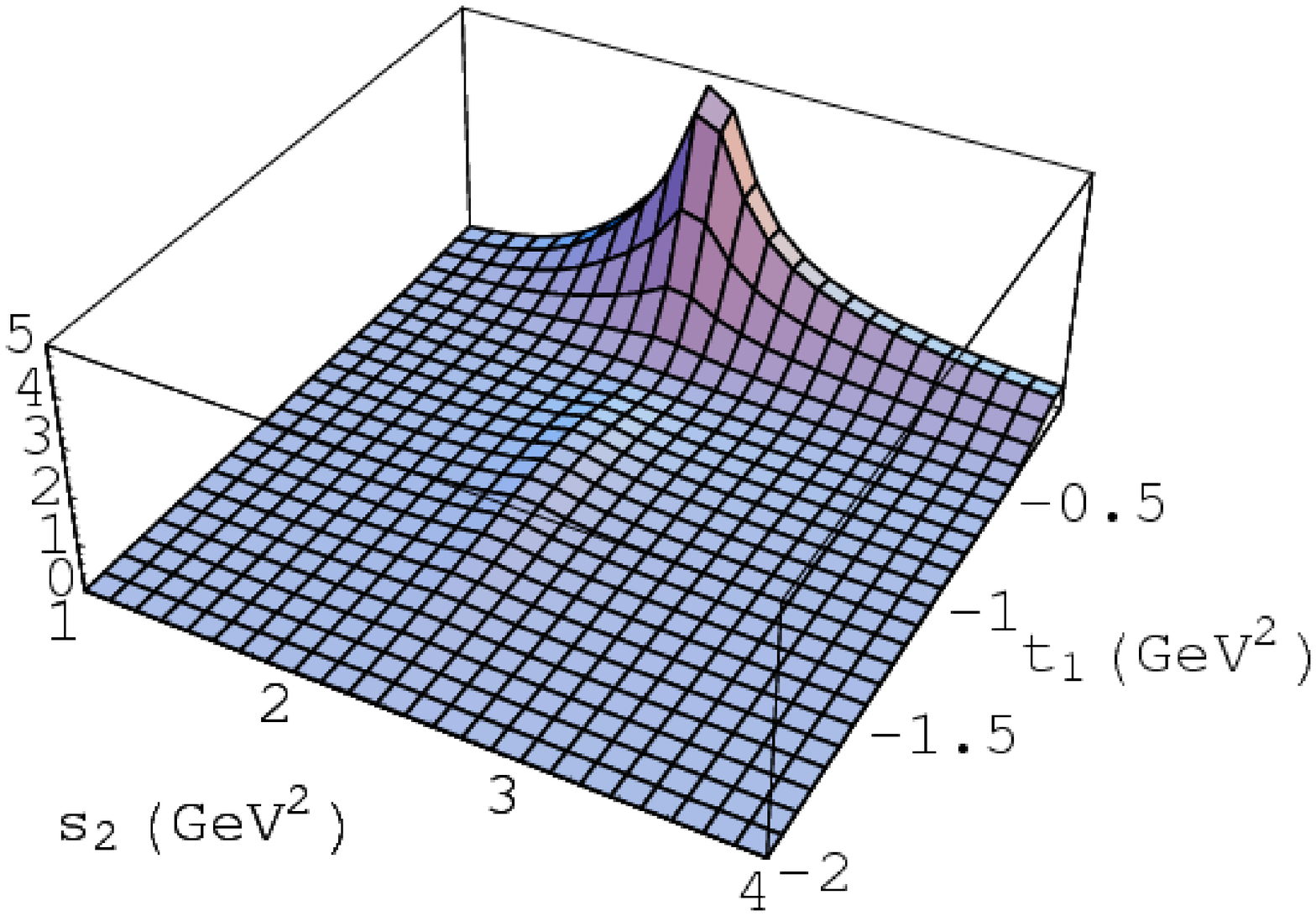,width=0.45\textwidth}
\caption{Double-differential cross sections $d\sigma/ds_2dt_1$ (in units of
nb/GeV$^4$) for $\gamma\gamma$ photoproduction off the proton. The same 
model as in Fig.2 for the soft set of $VNN$ coupling constants (Variant I). 
Left: $E_\gamma=\,$5 GeV. Right: 9 GeV.
\label{fig:fig3}}
\end{center}
\end{figure}

\begin{figure}
\vspace{5mm}
\begin{center}
\epsfig{figure=fig4a.eps,width=0.45\textwidth}\hspace{5mm}
\epsfig{figure=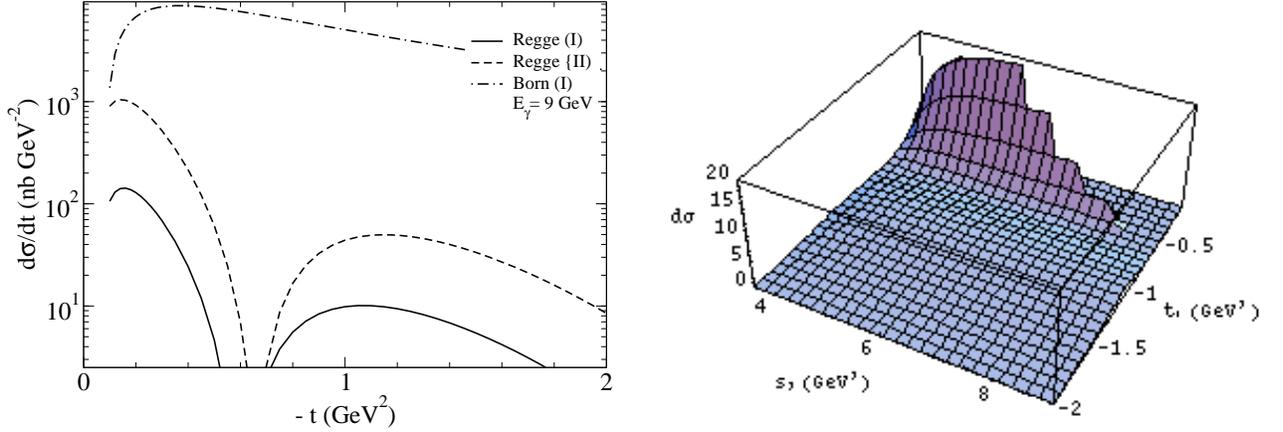,width=0.45\textwidth}
\caption{Differentioal (left panel) and double-differential (right panel) 
cross sections, $d\sigma/dt_1$ and $d\sigma/ds_2dt_1$, for $p\bar p$ 
photoproduction off the proton. The same model as in Figs.2-3 with the soft 
(Variant I) and hard (Variant II) sets of $VNN$ coupling constants. The same 
notations as in Figs.~2 and 3. $E_\gamma=\,$9 GeV.
\label{fig:fig4}}
\end{center}
\end{figure}

\end{document}